# A NEW APPROACH TO INDUCTION MOTOR DYNAMICS WITH PARAMETER PREDICTION

**Shayak Bhattacharjee**

Department of Physics,
Indian Institute of Technology Kanpur,
NH-91, Kalyanpur,
Kanpur – 208016,
Uttar Pradesh, India.

\* \* \* \* \*

# KEYWORDS

Induction motor   Electromagnetic theory   Iterative structure   Dynamic model

# ABSTRACT

In this work the dynamic model of an induction motor is derived from electromagnetic considerations. A motor model is constructed, and Maxwell's equations solved for it by an iterative method. The angular harmonics of the currents and magnetic fields are written as complex quantities analogous to the space phasor representation. Laplace transformation of the currents and fields, followed by a summation over all orders of the iteration leads to the transfer functions of the motor. An approximate case is considered first and is used to lead intuitively to the general case. The structure of our equations is identical to that of the existing models. In the process we provide theoretical estimates of the various inductances and resistances used in a traditional model. These estimates are in good agreement with the results of numerical evaluations.

\* \* \* \* \*

# I. INTRODUCTION

Induction motors are the backbone of industry due to their high performance and low maintenance requirements. The issue of controlling these motors gives rise to considerable interest in their dynamic behaviour. Below we present two popular motor models.

The d-q model [1]-[4] uses a two-phase motor in direct and quadrature axes. The rotor $α$ and $β$ axes are orthogonal and $α$ makes an angle $θ_r$ with the d axis of the stator. The terminal voltages of the stator and rotor windings are expressed as the sum of voltage drops across the resistances and the rate of change of the flux linkages. This latter term is expressed in terms of the self and mutual inductances and the rotor and stator currents. Under the assumption of uniform air-gap the self inductances are independent of the angular position and the mutual inductance between any pair of orthogonal windings is zero. These assumptions result in a system of differential equations with time-varying inductances. The transformation to obtain constant inductances is achieved by replacing the actual rotor with a fictitious one on the d and q axes. The fictitious rotor currents are related to the real ones through an idempotent matrix, which is applied on the system with time varying inductances, and finally referred to the stator. This results in the differential equations connecting the d and q components of rotor and stator current to the applied voltage. For the arbitrary frames model, the current and voltage vectors (d and q components) are multiplied by a matrix and substituted into the stator frames model to obtain the dynamical equations. This model is derived for a two phase motor, so it is necessary to transform from the actual three-phase to the hypothetical two phase machine. This is achieved by a matrix multiplication, known as Park transformation.

The space phasor model due to Kovacs and Racz [5] exploits the fact that sinusoidal distributions in space can be represented by complex variables. The space vector notation converts the voltage vector $[V_d, V_q]^T$ into the complex number $V_d + jV_q$ (where j is imaginary unit). The advantage is that the number of differential equations reduces from four to two. Furthermore, the shift from one reference frame to another is relatively easy. The equations are typically cast in terms of the flux linkages, though they may also be written in terms of the currents. The classic example of the latter formulation is in Takahashi and Noguchi's work [6] on DTC – the equation they have used for a 2-pole motor is

$$\begin{bmatrix} \mathbf{v_1} \\ \mathbf{0} \end{bmatrix} = \begin{bmatrix} R_1 + pL_{11} & pM \\ (p - j\dot{\theta}_m)M & R_2 + (p - j\dot{\theta}_m)L_{22} \end{bmatrix} \begin{bmatrix} \mathbf{i_1} \\ \mathbf{i_2} \end{bmatrix} \quad , \quad (1)$$

where $\mathbf{v_1}$ is the stator voltage vector, $R_1$ the stator resistance, p the differential operator, $L_{11}$ the stator self-inductance, M the mutual inductance, $\dot{\theta}_m$ the rotor angular velocity, $R_2$ the rotor resistance, $L_{22}$ the rotor self-inductance and $\mathbf{i_1}$ and $\mathbf{i_2}$ the stator and rotor currents. Recently, work by Holtz [7] has given special importance to the transfer function obtained from the dynamics as he has used them to plot complex signal flow graphs. These graphs have been used to understand the nature of the rotor-stator interaction inside the motor.

In this work we propose a different approach to motor dynamics. We derive it starting from Maxwell's equations for the motor. It is not possible to solve the equations directly so we develop an iterative method, Sec. II-1. The application of this method is shown in Sec. II-2. A limiting case is considered first so as to simplify the calculations involved. The answers of this restricted scenario can be interpreted in a physically meaningful way, which is extended to apply to the general case, II-3. Our equations are found to be identical in structure to (1) above. Equating the coefficients of corresponding terms in the two formalisms, we obtain theoretical formulae for the various resistances and inductances involved. In Sec. III we compare our findings with the results obtained from finite element analysis and demonstrate good agreement between the two.

# II. DERIVATION

Since the entire derivation is rather long and involved, we split it into various parts, doing a particular chain of calculations in each part.

## 1. MODELING AND ITERATIVE DEVELOPMENT

A schematic diagram of the induction motor is shown in Fig. 1. The cross section of the cylindrical motor is visible there. The outer part is the stator which has been shown to have a polarity of 2. The conductor bars in each labelled phase can be clearly seen. The inner part is the squirrel cage rotor where the bars are shorted by the end ring, visible in the figure. In this paper we will not deal with wound rotors. We define the various geometrical parameters in Table I.

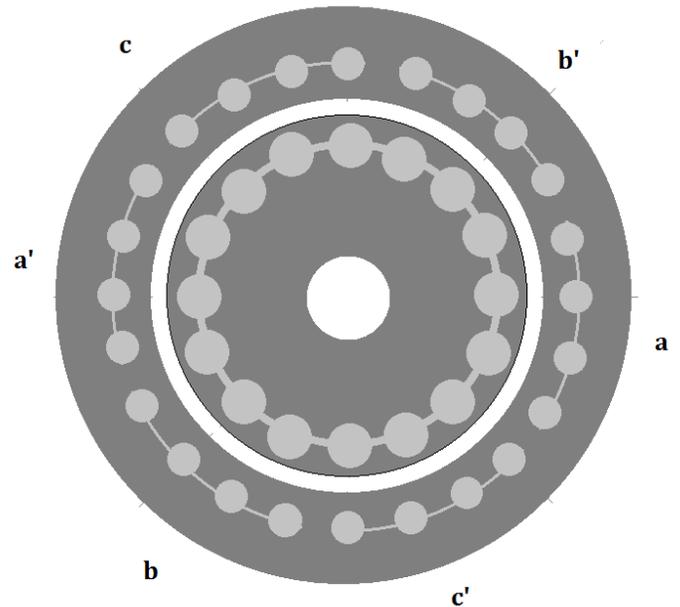

Figure 1 : The cross-sectional view of induction motor

| Parameter | Rotor | Stator |
|---|---|---|
| Height | h | h |
| Radius | r | R |
| Material conductivity | $\sigma$ | $\sigma'$ |
| Thickness | b | b' |
| Angular velocity (anticlockwise) | $\omega$ | NULL |
| Permeability of core | $\mu_c$ | $\mu_c$ |
| Permeability across air gap | $\mu_g$ | $\mu_g$ |
| Conductor separation factor | c | c' |

Table 1 : *Motor parameters involved in modeling. The polarity is 2n. Some parameters will be explained during the course of the analysis.*

The angular velocity ascribed to the stator in the above table shows that we will perform the entire analysis in the stator frame. A few geometrical observations lead to the mathematical model. The shape of the motor motivates the use of cylindrical co-ordinates $\rho,\theta,z$ with **z** along the motor axis. Since most motors are heavily elongated in structure, we assume the magnetic fields and currents to be uniform along **z**. Furthermore we assume that the currents flow exclusively along **z**. This amounts to neglecting end effects and rotor bar skew. For a **2n**-pole motor we note that the principal spatial harmonic of the fields and currents are expressible in the form **a cos(n$\theta$)+b sin(n$\theta$)** where $\theta$ is the azimuthal angle. Analogous to the space phasor representation, we write such a quantity as **a+jb** (where $j^2$=-1).

As regards the conductor bars, we are not going to use the circular bar geometry shown in Fig. 1. Rather, at first we will model the stator and rotor as continuous cylindrical shells of thickness **b** and **b'**. We will subsequently introduce a correction which will incorporate the discretized nature of the bars but will implicitly assume them to look like sectors of an annulus. An approximate prescription for generalizing to other bar shapes will be given later. Since the thicknesses **b** and **b'** are in general much less than the radii **r** and **R**, we work in terms of the *surface currents* (we may continue to use the word current for it, for brevity) flowing in the rotor and stator. This is a convenience as the magnetic field for a cylindrical shell carrying a sinusoidal surface current can be calculated easily.

We have written two permeabilities in the above table, so we explain the reason for it. Suppose some current is flowing through the stator. Then the magnetic field produced at a point within the stator core will carry a factor of $\mu_c$ which is the natural permeability of the core. When the same current creates a field in the rotor however, the permeability is reduced because of the air gap, and this reduced permeability will be denoted by $\mu_g$. We shall work in terms of the vector potential **A** such that $\nabla \times \vec{A} = \vec{B}$ where **B** is the magnetic field. Until this point, the surface currents, vector potentials and magnetic fields are all treated as vectors in real space. However, we realize that the only relevant components are the **z** components of the currents and potentials and the *ρ* component of the magnetic field. Because of this, we now treat them as spatial scalars and instead use the vector notation to denote the space phasor.

A direct solution of Maxwell's equations is hindered by our ignorance of the constitutive relation between electric fields and currents. In particular, the *σ* of **J**=*σ***E** may be a complex quantity. Our knowledge of the inductive character of the stator and rotor convinces us that it is indeed so. We note that at the frequencies involved, complex *σ*'s are not an inherent property of any material (they may become an inherent property at optical frequencies). Rather, they are a property of the way in which the conducting material is arranged, i.e. the specific current distribution. They arise from electromagnetic effects which occur within the concerned circuit, and can be considered as an abstraction which allows us to obtain the circuit's behaviour without analysing these effects in detail. Here we perform the detailed analysis to bypass the complex *σ*.

The analysis will consist of a step-by-step investigation of the currents and magnetic fields in the motor. Ultimately, the dynamics will emerge by combining all steps. In the zeroth step (it will soon become clear what the "step" exactly is) voltage is externally applied only on the stator. Let it produce a current as it would in a pure resistor. We make this assumption as right now we are concerned only about the material properties; any single step does not involve the complex *σ* which arises only after combining all steps. Now this current will create a magnetic field at the surface of both rotor and stator. If the applied voltage is not constant in time (which it generally isn't) the magnetic field will also change. By Faraday's law, electric fields will be induced in both the rotor and stator. These electric fields will give rise to currents in both elements, once more through real conductivities. Now these currents will create magnetic fields of their own, and again these fields will give rise to more currents, so the creation of fields and currents continues infinitely. Each "step" mentioned at the start of this paragraph is of course the creation of a new round of currents. Now by the principle of superposition, the resultant currents and fields will simply be the sum of all the contributing terms. Because of this structure we will construct series developments of the rotor and stator currents. The zeroth order currents (**K**$_{r0}$ for the rotor and **K**$_{s0}$ for the stator) will be the result of the applied voltage and will be proportional to the voltage through some factor dependent on the geometry of the motor and the materials used in it. **K**$_{r0}$ is obviously zero, and **K**$_{s0}$ will be denoted by the symbol **V**, a dimensional misnomer whose physical significance outweighs this minor contradiction. The first order currents Δ**K**$_{r1}$ and Δ**K**$_{s1}$ will follow from the effect of the zeroth order terms, the second from the first and so on. Summation of the series to infinity, if possible, will give an explicit formula for the dynamics of the currents. The problem has thus been separated into two components : (a) determination of the general terms Δ**K**$_{rk}$ and Δ**K**$_{sk}$ and (b) performing the infinite summation. The next subsection deals with the first of these objectives.

## 2. DETERMINATION OF THE GENERAL TERM

We note that the physical processes in going from the set [Δ**K**$_{rk}$, Δ**K**$_{sk}$]$^T$ to the set [Δ**K**$_{r(k+1)}$, Δ**K**$_{s(k+1)}$]$^T$ will be the same for all **k**, and so will the mathematical operator representing this transition. Hence it suffices to derive the structure of this operator for arbitrary **k**. Suppose we are given Δ**K**$_{rk}$ and Δ**K**$_{sk}$; we try to estimate Δ**K**$_{r(k+1)}$ and Δ**K**$_{s(k+1)}$. The estimation is done as per the phasor diagram in Fig. 2 which shows current, field and vector potential phasors in the **d**-**q** plane.

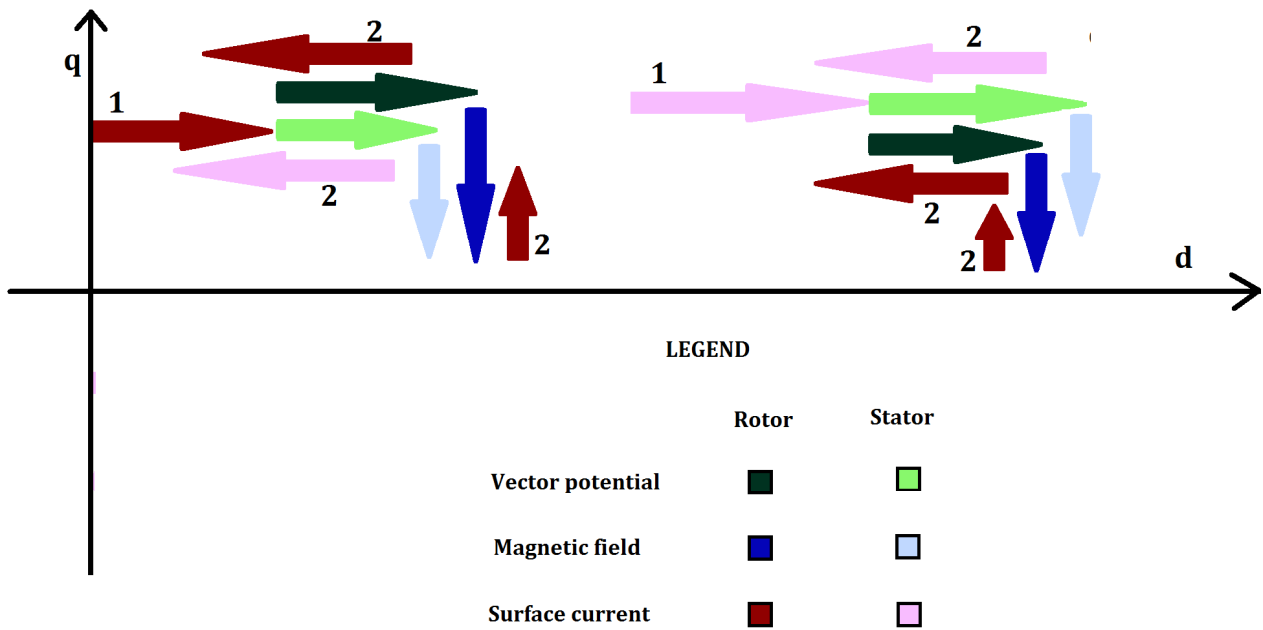

Figure 2 : *Currents, magnetic fields and vector potentials in the d-q plane.*

Here all rotor quantities are indicated in dark color and all stator quantities in light color. Red, green and blue stand for current, vector potential and magnetic field phasors respectively. The left panel shows the various phasors originating from one rotor current (dark red phasor marked 1). Under the magnetostatic approximation, valid so long as $\omega r$ is much less than the velocity of light, the vector potential due to this current is calculated by using the equation $\nabla^2 \vec{A} = -\mu_0 \vec{J}$ where **J** is the current density. Since the currents are confined to the stator and rotor surfaces, we get a Laplace equation. The boundary conditions are that the **A** is continuous across a current surface while the tangential component of **B** experiences a discontinuity proportional to the surface current. The formulae for the vector potential inside and outside a single cylinder (rotor) of polarity **2n** carrying a current vector **K** are given below.

$$\mathbf{A}_{in} = \frac{\mu_c \mathbf{K} r}{2n}\left(\frac{\rho}{r}\right)^n , \qquad (2a)$$

$$\mathbf{A}_{out} = \frac{\mu_g \mathbf{K} r}{2n}\left(\frac{r}{\rho}\right)^n . \qquad (2b)$$

It is seen from (2) that the vector potential phasor created at the rotor surface is parallel to the current phasor. So is the potential created at the stator surface, but its magnitude is smaller than the rotor surface potential. The differential form of Faraday's law, $\mathbf{E} = -\partial \mathbf{A}/\partial t$ gives the electric field. The fields are thus opposite in direction to the potentials. They are marked with a 2 to indicate that they are of the next order to the phasor marked 1. Now an additional component of rotor electric field arises from the fact that it is being dragged through a magnetic field. Alternatively, we must apply $\mathbf{J} = \sigma \mathbf{E}$ in the rotor frame, and frame transformation adds on a $\mathbf{v} \times \mathbf{B}$ term to the stator frame field. The magnetic fields are obtained as the curl (in real space) of the vector potentials and evaluate to

$$\mathbf{B}_{in} = -j\frac{\mu_c \mathbf{K}}{2}\left(\frac{\rho}{r}\right)^{n-1} , \qquad (3a)$$

$$\mathbf{B}_{out} = -j\frac{\mu_g \mathbf{K}}{2}\left(\frac{r}{\rho}\right)^{n+1} . \qquad (3b)$$

The magnetic field is thus orthogonal to the corresponding potential phasor and the resulting rotor electric field, also marked with a 2, is given by $-\omega r \mathbf{B}$. The current density **J** is obtained by multiplying **E** with the conductivity and the surface current is given by **J**b. This completes the development of the rotor current. The development of one stator current phasor is identical except for the numerical values of the coefficients and is shown in the right panel in Fig. 2.

As promised earlier we now refine the assumption that the rotor and stator are continuous cylindrical shells. In reality they consist of bars of wires with spaces between them. We assume that the shape of the bar is that of an annular sector so that each bar has a constant angular span. When the effect of the space is included, the currents will no longer be of the form $K\sin\theta$ (or $\cos\theta$) but will get multiplied by the Fourier expansion of the function which is unity in the angular intervals corresponding to the bars and zero in the inter-bar spaces. The first term of this expansion is easy; it is simply the ratio of the angular extension of each bar to that of each bar-plus-space unit. In other words it is like the "duty cycle" of the conductor in the cage circumference. This has been defined in Table 1 as the conductor separation factor. Incorporation of this factor will result in the potentials and fields created by the rotor getting multiplied by c, and those created by the stator by c'. Putting all this together yields the general transformation relation between successive rounds of current phasors, as follows.

$$\begin{bmatrix} \Delta \underline{K}_{r(k+1)} \\ \Delta \underline{K}_{s(k+1)} \end{bmatrix} = \begin{bmatrix} (c\mu_c \sigma rb/2n)(-p+jn\omega) & (r/R)^{n-1}(c\mu_g \sigma rb/2n)(-p+jn\omega) \\ (r/R)^{n+1}(c'\mu_g \sigma' Rb'/2n)(-p) & (c'\mu_c \sigma' Rb'/2n)(-p) \end{bmatrix} \begin{bmatrix} \Delta \underline{K}_{rk} \\ \Delta \underline{K}_{sk} \end{bmatrix}. \quad (4)$$

Letting **T** denote the transformation matrix in the above equation, we at once have the relation between the k[th] and zeroth order current phasors. Since the zeroth phasors are $[0,V]^T$ we have

$$\begin{bmatrix} \Delta \underline{K}_{rk} \\ \Delta \underline{K}_{sk} \end{bmatrix} = \vec{\vec{T}}^k \begin{bmatrix} 0 \\ V \end{bmatrix}. \quad (5)$$

This completes our objective of determining the general terms of the iterative development, and the only remaining task is that of

## 3. DOING THE INFINITE SUM

The form of (4) suggests the following variable definitions.

| New variable | Parameter combination | Physical significance |
|---|---|---|
| $\tau_r$ | $c\mu_c \sigma rb/2n$ | Rotor time constant |
| $\tau_s$ | $c'\mu_c \sigma' Rb'/2n$ | Stator time constant |
| $\delta_1$ | $(r/R)^{n-1}(\mu_g/\mu_c)$ | Incorporates mutual and leakage terms |
| $\delta_2$ | $(r/R)^{n+1}(\mu_g/\mu_c)$ | |

Table 2 : *Convenient variable definitions.*

The physical significance of the parameters mentioned above is not yet apparent but will become clear as the analysis proceeds. Since derivatives are involved and summation over repeated differentiations is tricky, we take the Laplace transform of (4), denoting the transform of **f**(t) by **f**(s). The summation over all k then yields the transfer function. The transformed (4) is

$$\begin{bmatrix} \Delta \underline{K}_{r(k+1)} \\ \Delta \underline{K}_{s(k+1)} \end{bmatrix} = \vec{\vec{T}} \begin{bmatrix} \Delta \underline{K}_{rk} \\ \Delta \underline{K}_{sk} \end{bmatrix} = \begin{bmatrix} \tau_r(-s+jn\omega) & \delta_1 \tau_r(-s+jn\omega) \\ \delta_2 \tau_s(-s) & \tau_s(-s) \end{bmatrix} \begin{bmatrix} \Delta \underline{K}_{rk} \\ \Delta \underline{K}_{sk} \end{bmatrix}. \quad (6)$$

The limiting case of $\delta_1, \delta_2 \ll 1$ is analysed first. It arises for motors with (a) wide air gap and (b) high polarity. In this case we construct two linear combinations of the rotor and stator currents such that **T** is replaced by a diagonal matrix. This calls for a determination of the eigenvalues and eigenvectors of its transpose and the smallness of the $\delta$'s is exploited in this step. The largest term of the stator current will be independent of $\delta$ (either 1 or 2) while the largest term of the rotor current will be first order in $\delta_1$ as it is driven by an interaction from the stator side. These considerations determine the precision to which the calculations must be carried out. The transfer functions for these current combinations are determined by summation over all correctional terms and are unscrambled to find the stator current transfer function

$$\underline{\mathbf{K}}_s = \frac{\underline{\mathbf{V}}}{1+\tau_s s} \quad , \tag{7}$$

and the rotor current transfer function

$$\underline{\mathbf{K}}_r = \frac{\delta_1(-\tau_r s + jn\tau_r \omega)\underline{\mathbf{V}}}{(1+\tau_s s)(1+\tau_r s - jn\tau_r \omega)} \quad . \tag{8}$$

Now this procedure is difficult to generalize to the case where there are no restrictions on $\delta$ so we attempt an interpretation in terms of the electromagnetic interaction going on inside the motor. The (1,1)th term of $\underline{\mathbf{T}}$ denotes the process by which a rotor current phasor produces the next order rotor current, hence it symbolises the rotor interaction with itself. The (2,2)th element likewise indicates the stator self-interaction and the off-diagonal terms represent the cross-interaction. In (7), the zeroth order stator current phasor is the applied $\underline{\mathbf{V}}$. The primary factor governing the stator currents is its interaction with itself – the rotor interacts with it through the $\delta_2$ term, which is very small. Hence the transfer function is the result of the stator interacting with itself an infinite number of times and producing terms like $\underline{\mathbf{V}}+(-\tau_s s)\underline{\mathbf{V}}+(-\tau_s s)^2\underline{\mathbf{V}}+.....$ a geometric series which quite obviously converges to the value given in (7). For the rotor, it might help to write (8) in a rearranged way :

$$\underline{\mathbf{K}}_r = \left\{\frac{1}{1+\tau_r(s-jn\omega)}\right\}\left\{\left[\delta_1\tau_r(-s+jn\omega)\right]\left[\frac{\underline{\mathbf{V}}}{1+\tau_s s}\right]\right\} \quad . \tag{9}$$

Considering the RHS of (9) we recognize the term in the first curly bracket as the result of an infinite number of rotor self-interactions. The second curly bracket which is essentially the zeroth order rotor current, has two parts (box brackets). The first box bracket is in fact the (1,2)th element of $\underline{\mathbf{T}}$ which gives the relation between a stator current and the next order rotor current. Hence the second curly bracket is like the rotor current produced by the stator current contained in the second box bracket. Now this last term is nothing but the *resultant* stator current, (7). This motivates the following observations, in which the restriction on the size of $\delta$ is lifted.

Let the resultant rotor current $\underline{\mathbf{K}}_r$ be known. Then it produces a zeroth order current phasor $\delta_2(-\tau_s s)\underline{\mathbf{K}}_r$ in the stator. The total zeroth order stator current is obtained by adding $\underline{\mathbf{V}}$ to this contribution. Now by including the *resultant* rotor current in the zeroth order stator term we have in effect taken into account all orders of the rotor-stator interaction. In other words we have effectively removed the rotor by incorporating its entire contribution into the initial stator term. Hence we now need to consider only the stator interaction with itself, leading to the transfer function

$$\underline{\mathbf{K}}_s = \frac{\underline{\mathbf{V}} - \delta_2\tau_s s\underline{\mathbf{K}}_r}{1+\tau_s s} \quad . \tag{10}$$

The identical procedure for the rotor with known stator current leads to the transfer function

$$\underline{\mathbf{K}}_r = \frac{\delta_1\tau_r(-s+jn\omega)\underline{\mathbf{K}}_s}{1+\tau_r(s-jn\omega)} \quad . \tag{11}$$

The corresponding dynamical equations are

$$\begin{bmatrix} 1+\tau_r(p-jn\omega) & \delta_1\tau_r(p-jn\omega) \\ \delta_2\tau_s p & 1+\tau_s p \end{bmatrix}\begin{bmatrix} \underline{\mathbf{K}}_r \\ \underline{\mathbf{K}}_s \end{bmatrix} = \begin{bmatrix} \mathbf{0} \\ \underline{\mathbf{V}} \end{bmatrix} \quad . \tag{12}$$

This is identical in structure to (1). The coefficients can be matched when we realize that n=1 for a 2-pole motor. Moreover, our equations have effectively normalized the resistances to unity as we consider surface currents on both the left and right hand sides. Conversion of these surface currents to voltages and currents through the necessary geometric factors will yield the resistances. It may appear that the coefficients on the off-diagonal terms are unequal in (12) whereas they are equal in (1). However this is an effect of not including the resistances. In particular, the first row of the matrix equation may be multiplied by any arbitrary constant as the rotor voltage on the RHS is zero. This constant can be chosen so as to equate the coefficients of the off-diagonal elements in the matrix.

We now make a brief note on the extension of our formalism to bars of arbitrary shape. If the bars are too deep, then of course the entire model will have to be reformulated. If the depth is not a problem but only the shape is not close to that of the annular sector which we have assumed, then we suggest obtaining equivalent values of c and b (and c' and b') from the area of the bars. If our rotor (identical comments hold for the stator so we consider only the rotor) has x bars, then the area of each as per our model is $a_{bar} = rb\frac{2\pi c}{x}$. For a real motor x, r and $a_{bar}$ will be directly measurable and we can take the effective product $c_{eff} b_{eff} = \frac{a_{bar} x}{2\pi r}$.

It is quite easy to generalize the model to arbitrary frames. Let us view the motor from a reference frame rotating anticlockwise at angular velocity $\beta$. Then the apparent angular velocity of the rotor will become $\omega$-$\beta$ and the apparent angular velocity of the stator will be $-\beta$. The motional emf effect, as discussed in the analysis of Fig. 1, will now contribute to the stator also. The reformulated (6), where the arbitrary frame voltage and current vectors are denoted by a superscript **e**, will become

$$\begin{bmatrix} \Delta\underline{\mathbf{K}}^e_{r(k+1)} \\ \Delta\underline{\mathbf{K}}^e_{s(k+1)} \end{bmatrix} = \begin{bmatrix} \tau_r(-s+jn(\omega-\beta)) & \delta_1\tau_r(-s+jn(\omega-\beta)) \\ \delta_2\tau_s(-s-jn\beta) & \tau_s(-s-jn\beta) \end{bmatrix} \begin{bmatrix} \Delta\underline{\mathbf{K}}^e_{rk} \\ \Delta\underline{\mathbf{K}}^e_{sk} \end{bmatrix} . \quad (13)$$

The steps leading to (12) from (6) can be retraced to obtain the arbitrary frame dynamics as

$$\begin{bmatrix} 1+\tau_r(p-jn(\omega-\beta)) & \delta_1\tau_r(p-jn(\omega-\beta)) \\ \delta_2\tau_s(p+jn\beta) & 1+\tau_s(p+jn\beta) \end{bmatrix} \begin{bmatrix} \mathbf{K}^e_r \\ \mathbf{K}^e_s \end{bmatrix} = \begin{bmatrix} \mathbf{0} \\ \mathbf{V}^e \end{bmatrix} . \quad (14)$$

We must now write the mechanical equation for the motor i.e. the one describing the time evolution of $\omega$. The first thing to take note of is that an integration over $\theta$ is required where the quantity to be integrated is the product of two angular distributions – the rotor current and the magnetic field at the rotor surface. Using the standard identities for integrals of products of trigonometric functions, we get that when two distributions are multiplied and integrated, the product of the respective *d* components will have a nonzero contribution to the result as will the product of the respective *q* components. The cross terms, featuring *d* of one phasor and *q* of the other, will however evaluate to zero. Hence for arbitrary distributions *X* and *Y*,

$$\int_0^{2\pi} (X_d \cos n\theta + X_q \sin n\theta)(Y_d \cos n\theta + Y_q \sin n\theta) \mathrm{d}\theta = \pi(X_d Y_d + X_q Y_q) . \quad (15)$$

The quantity of the RHS of (503) looks very much like the dot product of the phasors **X** and **Y**, hence the integral of the product of two distributions can be written as the dot product of their representative phasors. The rotor torque is thus proportional to the dot product of the rotor current phasor and the magnetic field phasor at the surface of the rotor. The rotor current $\mathbf{K}_r$ is determined from (460) and the field will have two contributions – one from the rotor itself and the other from the stator. Equation (433) implies that the rotor field phasor is orthogonal to $\mathbf{K}_r$, hence it will not contribute to the dot product. The only contribution will be from $\mathbf{K}_s$, and this will be obtained from (437). Hence the key term in the expression for the torque $\Gamma$ will be

$$\Gamma \propto \mathbf{K}_r \cdot \left( -j\delta_1 \frac{\mu_c}{2} \mathbf{K}_s \right) \quad . \tag{16}$$

From this point on, it is a matter of careful book-keeping to take all the factors of $r$, $h$ etc. into account. The second term of the dot product in (504) gives the magnetic field and the first term is the surface current. Multiplication of this by $rd\theta$ gives the infinitesimal current – the $d\theta$ is already accounted for but we have collected a factor of $r$. Then the infinitesimal force on the conductor is obtained from $I\times\mathbf{B}$ (no space phasors here) which throws in a factor of $h$. And torque involves $\mathbf{r}\times\mathbf{F}$ so yet another factor of $r$ is accumulated. Finally we have to include the $\pi$ from the RHS of (503). This will give us the magnitude, so now let us obtain the direction. By default, $\mathbf{K}_r$ and $\mathbf{K}_s$ are in the $+z$ direction, and $\mathbf{B}$ in $+\rho$. Hence, if $\mathbf{K}_r$ and $\mathbf{K}_s$ are both positive, the force will be $\hat{\mathbf{z}}\times\hat{\boldsymbol{\rho}}=+\hat{\boldsymbol{\theta}}$ and the torque will be $\hat{\boldsymbol{\rho}}\times\hat{\boldsymbol{\theta}}=+\hat{\mathbf{z}}$. Combining all this we have

$$\Gamma_z = \frac{\pi\mu_c \delta_1 r^2 h}{2} \mathbf{K}_r \cdot (-j\mathbf{K}_s) \quad . \tag{17}$$

This formula gives the output torque of the given motor.

## III. DISCUSSION AND CONCLUDING REMARKS

This paper obtains the motor dynamics starting from first principles. In addition to the dynamic structure we have provided theoretical estimates of the various resistances and inductances involved in the modeling. Prediction of these parameters is difficult [9] and they are generally determined from experiments on the motor, or from numerical techniques. Let us compare the results derived here with the predictions of the accurate Finite Element Analysis (FEA) method. The classic paper is by Williamson [10] where they have, among other things, plotted the flux density inside the rotor. Their flux density plot (Fig. 6) is reproduced below as Fig. 3. Our model of the 2-pole motor treats it as a cylinder with surface current $K\sin\theta$ and it is well known that such a distribution produces a uniform field inside. This is very close to the field obtained in [10]. Yet another plot of the magnetic flux, this time outside the motor, can be found in [11]. Like [10], FEA is used here. The motor is 4-pole and the magnetic field should be like that of a cylinder with surface current $K\sin 2\theta$ which is the description of the motor as per our model. The similarity between their figure 2 (reproduced as Fig. 4) and the quadrupole field, (3) with n=2 is readily apparent. A combination of analytical and numerical methods may be found in [12]. There the time constant can be obtained by dividing their equation (14)

$$L_b = \mu_0 l_b \lambda_{bar} \qquad (14) \text{ of } [12]$$

by their equation (8)

$$R_b = \frac{\rho_b l_b}{A_b} \qquad (8) \text{ of } [12]$$

to get a result proportional to the conductivity, permeability and the cross sectional area of the bar [see comment after (12)], and independent of the bar length. This is in direct agreement with our predictions, which in addition specify a numerical value of the indeterminate $\lambda_{bar}$. Our purely analytical approach has thus yielded ready-to-use estimates of the various motor parameters which are in good agreement with highly sophisticated simulations.

Our analysis can also be used for a theoretical study of parameter variation during operation. The main cause of such variation is the change in temperature of the motor during running. Knowledge of the temperature dependence of the various conductivities and permeabilities will allow the instantaneous inductances and resistances to be predicted by measuring the motor temperature. A temperature feedback loop may then be installed in a variable frequency drive [13] to take care of this variation. Our work will also enable a study of secondary effects such as skin effect in the rotor conductors or hysteresis and saturation in the core. Studies of these physical phenomena can be drawn on to cause suitable modifications of the various terms in (4). Let us elaborate a bit on skin effect. As the slip frequency increases, the effect becomes more and more prominent. The resistance will tend to increase as the effective cross-sectional area of the bar reduces. Simultaneously, the time constant, which is proportional to the area, should reduce. This is exactly what is found in [14]. Their Figure 3 clearly shows the increasing resistance and the reducing time constant with increase in slip.

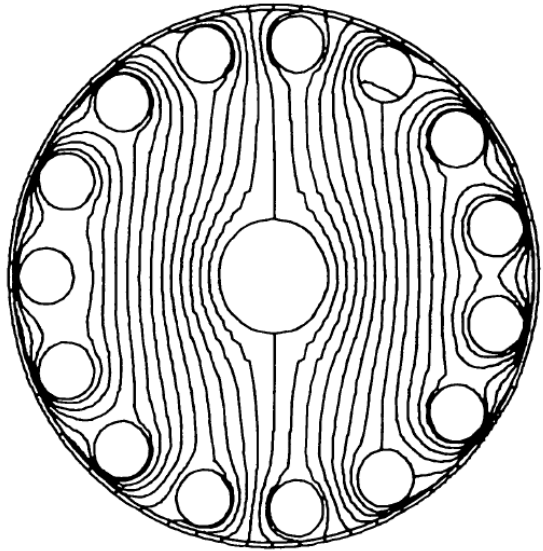

Figure 3 : *Flux distribution of 2-pole motor.*

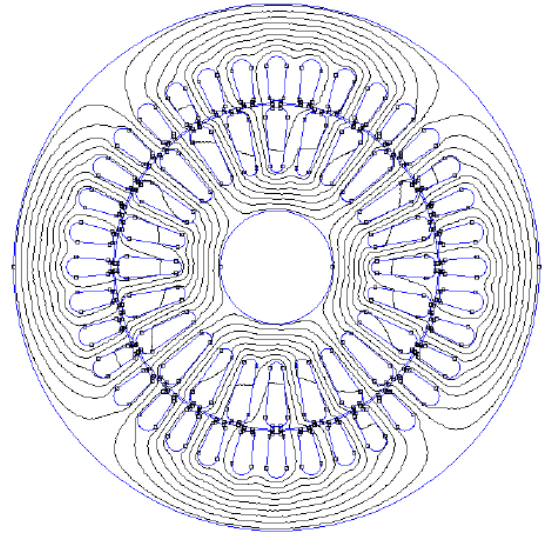

Figure 4 : *Flux distribution of 4-pole motor.*

Our work also helps in understanding the electromagnetic interaction going on inside the motor. Such a study has been undertaken by Holtz [15] who has started from a traditionally derived dynamic model (space phasor) and interpreted it in terms of the magnetic energy propagation inside the machine. Our process is in a sense the reverse – starting from the magnetic behaviour we have obtained the dynamic model. Such a derivation process has also enabled us to examine the limit $\delta_{1,2} \ll 1$ which is valid in motors of high polarity. In such a case, the dynamical equations are naturally uncoupled, which is a considerable simplification over the general case.

Though carried out for the principal harmonic, our method is applicable to all spatial harmonics of a particular motor. The values of the various parameters will be different for each harmonic, as per Table 2, but the dynamical structure remains unchanged. A detailed Fourier analysis of the applied voltage waveform can be carried out, including many spatial harmonics of the fields and currents. (12) or (14) can then be applied on each harmonic for a particularly accurate transient analysis. These relations can be used in high performing vector controlled pulse width modulator (PWM) drives. Finally, we note that our analysis is applicable not just to induction motors (of all phases – the phase did not enter the calculation anywhere) but to all kinds of electrical machines. The basic tenets of electromagnetism, on which our entire structure rests, are the same for induction, synchronous and dc motors, generators and transformers. Hence our method is universal in scope and can be profitably employed for an insightful study of all the electrical workhorses of modern industry.

\* \* \* \* \*

## ACKNOWLEDGEMENT

I am grateful to KVPY, Government of India, for a generous Fellowship.